# Optimization of Cellular Resources Evading Intra and Inter Tier Interference in Femto cells Equipped Macro cell Networks

Niraj Shakhakarmi

Department of Electrical & Computer Engineering, Prairie View A&M University (Texas A&M University System)
Prairie View, Houston, Texas, 77446, USA

**Abstract**
Cellular network resources are essential to be optimized in Femto cells equipped macro cell networks. This is achieved by increasing the cellular coverage and channel capacity, and reducing power usage and interference between femto cells and macro cells. In this paper, the optimization approach for cellular resources with installed femto cells in macro cell networks has been addressed by deploying smart antennas applications and effect power adaptation method which significantly optimize the cellular coverage, channel capacity, power usage, and intra and inter tier interference. The simulation results also illustrate the outstanding performance of this optimization methodology.

## 1. Introduction

Femto cells are the home base stations for better indoor voice and data coverage which quickly access the backbone infrastructure via internet. Femto cells provide better coverage and capacity, improved macrocell reliability, cost benefits and reduced subscriber turnover. The major technical perspective of femto cell is the capacity which can be optimized by increasing signal strength and mitigating interference from neighbouring macro cell. Besides this, femto cells also imply the efficient allocation of precious power and frequencies under the provision of Quality of Service (QoS) over the backhaul.

The research challenges in femot cells are broadband femto cells, voice femto cells and network infrastructure [1]. The physical and medium access layers in broadband femto cells addresses intra and cross tier interference, timing and synchronization, backhaul provision for acceptable QoS. Similarly, the physical and medium access layers in voice femto cells addresses cross tier interference, open or close access, open access handoff, outside coverage tracking emergency-911 and poor coverage to near close access, open access handoff, outside coverage tracking emergency-911 and poor coverage to nearby macro cells user. Network infrastructure includes packet switched over IP, IMS/SIP, RAN-gateway based unlicensed mobile access. But, the goal of efficient femto cells architecture leads the further research trends to MIMO femto cells and interference management via frequency and time hopping, directional antennas and adaptive power control.

Femto cells are the short transmission power base stations with high coverage and capacity for small isolated areas with insufficient or no macro cell coverage. Femto cells provide any existing internet access to cellular subscriber to access in the cellular communication networks. When the operating frequency of femto cell base station (BS) is same to macro cell BS, these co-channel cells create the cross tier interference between them [2]. The cross tier interference includes both femto cell to macro cell and macro cell to femto cell interference in downlink and uplink, which is the major issue to be resolved for the successful implementation of femto cell networks. The major reason is that it causes to both the macro cell and femto cell users suffer by dead zone in femto cell environment which degrades the link capacity and coverage. On uplink, dead zone occurs when cell edge macro cell user transmitting at maximum power causes deplorable interference to nearby femto cells. On downlink, dead zone occurs when the macro cell users have interference from close by femto cell transmission where they endure higher path loss [3-4].

Cross Tier Interference between femto cells and macro cells can be mitigated by optimal allocation of spectrum, outage constraint analysis, orthogonal sub-carriers, multi antenna and antenna sectorization. The optimal decentralized spectrum allocation policy in terms of Area Spectral Efficiency (ASE) is subjected to a sensible Quality of Service requirement, which guarantees that both macro cell and femto cell users attain at least a prescribed data rate [5-7]. This scheme is related to QoS requirement, hotspot density and the co-channel interference from the macro cell and surrounding femto cells [8-9].

Furthermore, accurate characterization of the uplink outage probability taking cross-tier power control, path-loss and shadowing can be used as the operating contour for the combinations of the average macro cell users and femto cell BS per cell-site that meet a target outage constraint [3]. The cross tier interference can also be eliminated as femto cell MSs and neighboring macro MSs use orthogonal subcariers to avoid mutual interference [4].

To maintain the orthogonality, the Macro MS should synchronize with both neighbor femto cell BSs and the macro cell BS. Another approach is equipping the macro cell and femto cells with multiple antennas enhance robustness against the near-far problem which derives the maximum number of simultaneously transmitting multiple antenna femto cells meeting a per-tier outage probability constraint [3-5]. All of above strategies need to be optimized further more in terms of outage and capacity by avoiding cross tier interference between macro cells and femto cells which is addressed in this paper and better illustrated in section-2 with novel results.

## 2. Problem & Proposed Solution

The problem is to analyze and compute the cellular outage, capacity and the impact of power control error on cellular capacity deploying DS-CDMA. Furthermore, the cross tier interference of Femto cells in DS-CDMA based Macro cell Networks is analyzed and addressed by the optimal solution of the integrated approach of the shared spectrum, sectored antenna and cognitive power control.

The proposed solution includes the following aspects:

- Cellular Outage Capacity
- Effects of Power Control Error on Cell Capacity
- Avoidance of Cross Tier Interference of Femto cells in Macro cell Network
- Simulation and Performance Evaluation of Femto cells in Macro cell Network

2.1 Cellular Outage Capacity

The desired or threshold signal to co-channel interference ratio (SIR) cannot meet by DS-CDMA cellular subscribers. This phenomenon is called outage and the outage probability ($\chi$) below threshold SIR can be defined as:

$$\chi = Pr[SIR < \gamma_o], \gamma_o - T_{SIR} \quad [1]$$

Considering equal power for all transmitters,

$$SIR = \left( \frac{G_o (1/r)^n}{\sum_{k=1}^{M} X_k G_k (1/D_k)^n} \right) \quad [2]$$

$$SIR = \frac{G_o}{G_I} \left( \frac{D}{r} \right)^n \quad [3]$$

$$G_I = \sum_{k=1}^{6} X_k G_k \quad [4]$$

where, $G_I$ = Composite fading;
$G_k$ = Log normally distributed with mean

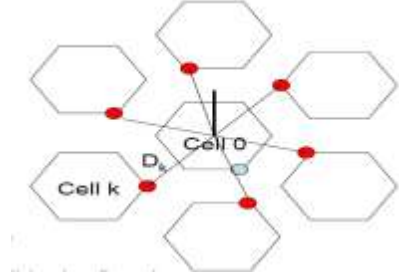

Fig. 1 Cellular Structure

The outage probability at the distance r is given by:

$$\chi = 1 - Q\left( \frac{(\gamma_o)_{db} - 10n\log(\partial) + M_I}{\sigma_{Tot}} \right) \quad [5]$$

$$\chi = Pr[10\log G_{Tot} \leq (\gamma_o)_{db}] - 10n\log(\partial) \quad [6]$$

$$\chi = \int_{r=0}^{R} 1 - Q\left( \frac{(\gamma_o)_{db} - 10n\log(\partial) + M_I}{\sigma_{Tot}} \right) p(r) dr \quad [7]$$

Using $Q(x) = \frac{1}{\sqrt{2*\pi}} \int e^{-t^2}/_2 \, dt$ and integrating as well solving above equation for $d>d_o$ and at cell boundary ($d=d_o$) yield as follows which can be further reduced considering parameters.

$$\Pr(a) = \left( e^{-a^2}/_2 - e^{-1/2} \right) \left\{ 2*\pi * e^{(-18-10*K*\log(a)+e(\sigma^2/2)^2/2)}/_{\sigma^2} - \sqrt{2*\pi} \right\} \quad [8]$$

where, a=d/$d_o$ and R=cell boundary

$$\Pr(a) = \left( e^{-2*R^2*\pi^2}/_2 - \pi/6 \right) \left\{ 2*\pi * e^{(-18-10*K*\log(a)+e(\sigma^2/2)^2/2)}/_{\sigma^2} - \sqrt{2*\pi} \right\} \quad [9]$$

The simulation is conducted for the outage capacity considering different locations of cell users $d_o$ inside the cellular coverage of d=500m from Base station such that $d>d_o$ and deploying exponent factor k=4 and 5 respectively as shown in Fig. 2-3.

The probability of fluctuation over desired or threshold signal to co-channel interference ratio (SIR) is the outage capacity of cell. The outage capacity (probability of S/I

<18 db) increases with respect to increasing distance of user (d/d$_o$) from the base station to the cell boundary. The channel statistics as exponential factor k is varied, the affect of k is found significant as the outage probability quickly optimised to unity at k=5 but it is greatly suppressed by other factors such as distance between user and base station as well as co-channel interference from the first tier. From the statistics of the simulation, it can be concluded that the outage capacity varies with respect to the position of the user rather than the channel statistics. As the user move away from the base station towards cell boundary with higher value of k, S/I decreases drastically and the S/I is the weakest instant of the desired value at the cell boundary. This means that the outage capacity of cell is the worst case at the cell boundary.

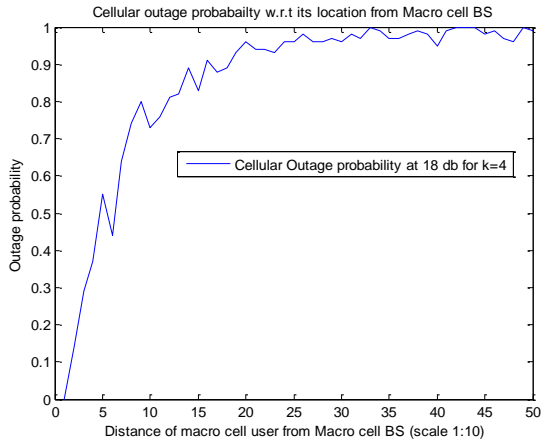

Fig. 2 Cellular Outage at $T_{SIR}$=18 db, k=4

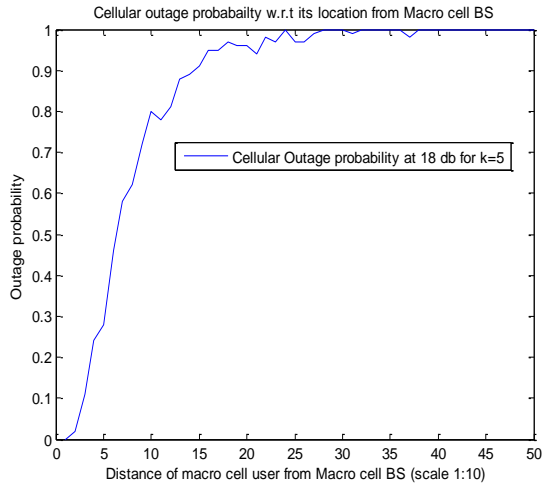

Fig. 3 Cellular Outage at $T_{SIR}$=18 db, k=5

Using an exact cell geometry layout for a seven-cell cluster as shown in Fig. 1, with the mobile unit at the cell boundary, the mobile is a distance D – R from the two nearest co-channel interfering cells and is exactly D + R/2, D, D – R/2, and D + R from the other interfering cells in the first tier, S/I is given as follows;

$$\frac{S}{I} = \frac{1}{2*(Q-1)^{-4} + 2*(Q+1)^{-4} + 2*Q^{-4}} \quad [10]$$

$$\frac{S}{I} = \frac{R^{-n}}{\sum_{i=1}^{io} D_i^{-n}} \quad [11]$$

For N = 7, the co-channel reuse ratio Q is 4.6, and the worst case S/I is approximated as 49.56 (17 dB) using equation-1 whereas an exact solution using equation-11 yields 17.8 dB. Hence, for a seven-cell cluster, the S/I ratio is slightly less than 18 dB for the worst case. To design the cellular system for proper performance in the worst case, it would be necessary to increase N to the next largest size. In other words, the co-channel interference determines link performance, which margins the frequency reuse plan, the overall capacity of cellular systems and it can be properly mitigated by physically separating co-channel cells by the optimal minimum distance.

## 2.2 Effects of Power Control Error on Cell Capacity

The DS-CDMA cell capacity depends upon the processing gain, bit energy to interference ratio, source activity factor, frequency reuse efficiency and number of sectors. The cell capacity with the capacity degradation factor ($C_d$) due to imperfect power control is given by;

$$N_c = 1 + (C_d * n_f) * \left\{ \left[ (Q*G_p) / E_{b_b}/I_o \right] - (P_n/S) \right\} / S_f \quad [12]$$

where, Cell Sectors(Q) = usually 3 for DS-CDMA

Processing Gain ($G_p$) =256

Frequency Reuse Efficiency Factor ($n_f$) =1/$K_f$

Imperfect Power Control ($C_d$ or σ)= 1 dB, 2 dB, 3dB

Bit energy to interference ratio ($E_b/I_o$)< 7 db or 8.45

Source Activity Factor ($S_f$ or α) = 1 %, 5%.

Signal to Background Noise Ratio (S/$P_n$) = 26 db

Similarly, the cell capacity with perfect power control is given as;

$$N_c = 1 + n_f * \left\{ \left[ G_p / E_{b_b}/I_o \right] - (P_n/S) \right\} \quad [13]$$

From the simulation results, the cell capacity decreases with respect to Power Control Error ($C_d$). At Source Activity Factor ($S_f$)=5% , the cell capacity is 4300, 2900, 2600 channels at the Power Control Error ($C_d$)=1db, 2db and 3db. Thus, Power Control Error ($C_d$) causes to

Multiple Access Interference (MAI) and decreases the Cell Capacity.

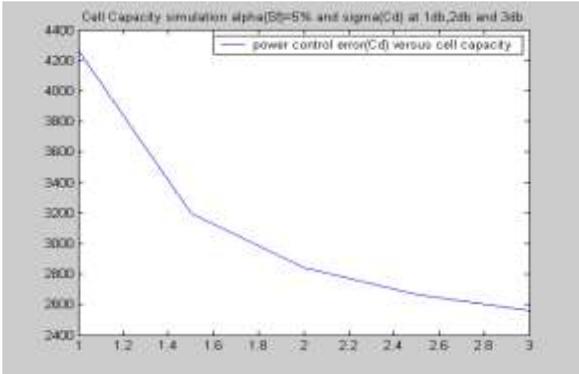

Fig. 4 Cell Capacity versus Imperfect Power Control at $S_f$ =5%

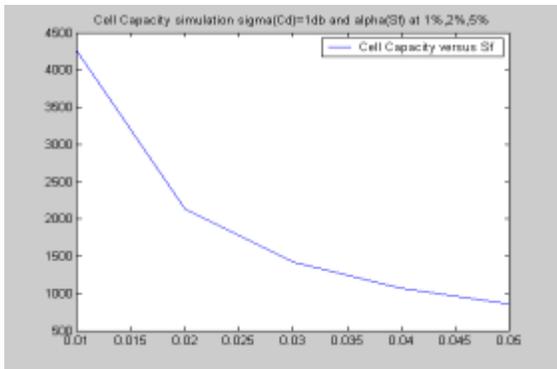

Fig. 5 Cell Capacity versus Source activity factor at $C_d$=1db

From the simulation results, the cell capacity decreases with respect to increment in Source Activity Factor ($S_f$). At Power control Error =1db, the cell capacity is 4200, 2300, 900 channels at the source activity factor ($S_f$)=1%, 2% and 5%. Thus, source activity factor increases Quality of service (QoS) but drastically decreases the processing gain which results the reduction in cell capacity.

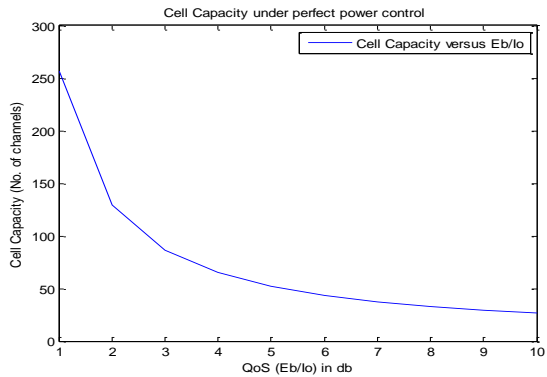

Fig. 6 QoS versus Cell Capacity

From the simulation results, the cell capacity decreases with respect to increment in Quality of service ($E_b/I_o$) at the perfect power control. Under perfect power control, the cell capacity is 260, 70, 40 channels at QoS of $E_b/I_o$=1db, 5db and 10db. This represents the trade off between the Cell Capacity and QoS as the cell capacity decreases with respect to increasing QoS under perfect power control. Thus, the DS-CDMA cell capacity decreases with increasing power control error, source activity factor with imperfect power control and QoS under perfect power control.

### 2.3 Avoidance of the Cross Tier Interference between Femto cell and Macro cell Networks

DS-CDMA macro cell networks provide significant power control for compensation of path-loss, shadowing and fading, which yields the uniform coverage in the absence of femto cells. But, the deployment of femto cells in macro cell develops dead-zones as well as non-uniform coverage. Basically, neighboring macro cell user using higher signal power generate interference to femto cell user's uplink as shown in Fig.7. Similarly, femto cell users use higher signal power as expressed in equation-15 for downlink that generates interference from femto cell BS to macro cell user as shown in Fig.8. In both case, one having higher signal power dominates other having lower signal power which is the cross-tier interference between macro cell and femto cell. On the other hand, when femto cells are deployed at the macro cell edge then closer macro cell users at edge boundary are interrupted by nearby femto cell transmissions in downlink forming dead zone. Similar dead zone incident happens to femto cell when macro cell users deploy higher signal in uplink at the edge boundary. This suffers higher path-loss to macro cell user and femto cell as shown in Fig. 9-10. Furthermore, the macro cell BS also get interference from the femto cells located in the other second tier as well as the co-channel macro cells.

The downlink power for femto cell BS is computed as follows;

$$P_{femto} = min\left(P_{macro} + G(\theta) - L_{macro(d)} + L_{femto(r)}, P_{max}\right) \quad [14]$$

where, $P_{macro}$: Tx power of macro cell BS

$G(\theta)$ : Antenna gain

$L_{macro}(d)$: Path loss between macro cell BS and Femtocell BS.

$P_{macro}+G(\theta) -L_{macro}(d)$: Received macro cell power by femto-user.

$L_{femto}(r)$: Path loss between femto BS and femto user.

$P_{femto}$: is about a few µW at macro cell edge to 125 mW macrocell center.

The uplink power for femto cell BS considering one active femtocell user in the femtocell, is computed as follows;

$$P_{UE} = min\left( \frac{P_{interference,max}}{n_{femto}} + L_{macro,measured}, P_{UE,max} \right)$$ [15]

where, $P_{interference,max}$: max. allowed interference from all the femto users to the macro-BS.

$n_{femto}$: number of femto cells in the considered sector.

$P_{interference,max}/n_{femto}$: allowed interference from the femto users in a femtocell.

$L_{macro,measured}$: Path loss between femto user and macrocell BS.

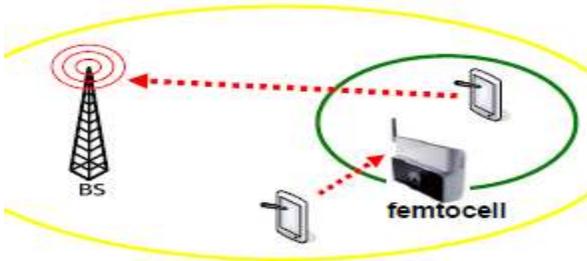

Fig. 7 Uplink Interference

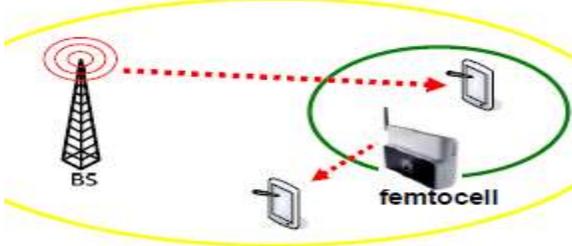

Fig. 8 Downlink Interference

An optimal technique for avoidance of cross tier interference is the integrated approach of the shared spectrum, sectored antenna and cognitive power control. First of all, the bandwidth spectrum is shared between femto cell and macro cell BS with proper assignment so that there will be always proportionately traffic balanced condition in spectrum utilization and it mitigates the hop spot density and cross tier interference. When the hot spot density is created among macro cell users then macro cell users can use the available spectrum of the nearest neighboring femto cell networks. Similarly, femto cell users can also use the available spectrum of macro cell BS within the range of femto cell networks to provide better QoS in the congested femto cell traffic, especially at the edge between co-channel macro cells. This is the logical way for sharing spectrum holes between macro cell and femto cells to regulate the optimal traffic load in co-operation.

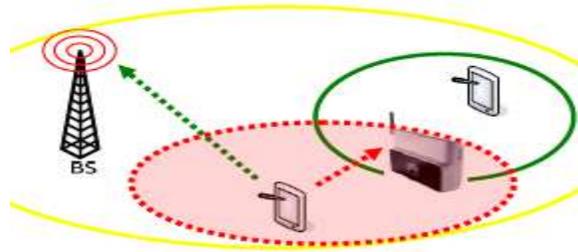

Fig. 9 Dead Zone in Uplink

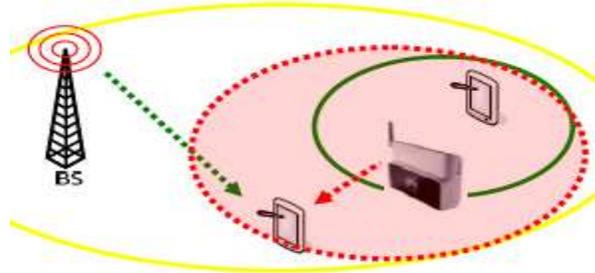

Fig. 10 Dead Zone in Downlink

As the spectrum is shared between femto cell and macro cell, then directional antennas are best to achieve sectorized antenna reception in both the macro cell and femto cell BS, with antenna alignment angle θ and sector width equaling $2\pi/N$ sector. Antenna sectoring is a common feature at the macro cell BS in real-time cellular systems, which is also recommended at femto cell BS. The major cause is that the cross-tier interference caused by some nearby macro cell users can lead to unacceptable outage performance over the femto cells uplink which could be nullified by adaptive beam forming to increase SNR in particular sectored area as shown in Fig.11 and provides good cell capacity and coverage. This is further optimized by deploying multi-sectored adaptive beam forming using the coded orthogonal frequency division multiplexing. In other words, the macro cell outage and femto cell outage computed in equation-16 and equation-17 caused by cross tier interference are significantly decreased by multi-sectored antenna deployment at both macro cell BS and femto cell BS.

The macrocell outage is computed as follows;

$$\Pr\left[\frac{GP_r^c}{I_{c,in}+I_{c,out}+I_{c,f}} \le \gamma_c\right] \le \varphi_c \quad [16]$$

where,

$I_{c,in}$: In-cell macro-cell interference assuming Poission

$I_{c,out}$: Out-of-cell macro-cell interference assuming Gaussian

$I_{c,f}$: Femto cell cross-tier interference with stability exponent δ=2/α. This is significantly reduced by a factor 'n' deploying 'n' number of sectored beam at femto cell BS because the interference is reduced by number of sectored beam.

The femto cell outage is computed as follows;

$$\Pr\left[\frac{GP_r^f}{I_{f,in}+I_{f,out}+I_{f,f}} \le \gamma_f\right] \le \varphi_f \quad [17]$$

where,

$I_{f,in}$: Intra-tier interference from active femto cell users within same femto cell assuming Poission

$I_{f,out}$: Inter-tier interference from femto cell BS from Other tier macro-cell assuming Gaussian

$I_{f,f}$: Femto cell interference from other femto cell BS within same macro cell.

Finally, cognitive power control strategy vary the femto cell received power target depending on its location to remove the possible cross tier interference at the junction of directional antennas and edge of femto cell and macro cell networks to avoid the severe signal jamming or deadlock. As the distance is increased between femto cell to macro cell users at the edges of femto cell, the possible interference to macro cell users is decreased by reducing the femto cell transmit power to avoid the possible dead zones. Thus, this approach avoids the entire cross tier interference and prevent from dead zones formation between femto cell and macro cell BSs. This is the best methodology for optimal avoidance of cross tier interference because it avoids dead zone and deals even in hotspot density with cognitive power control, resulting the good capacity, coverage and reliability rather than other approach.

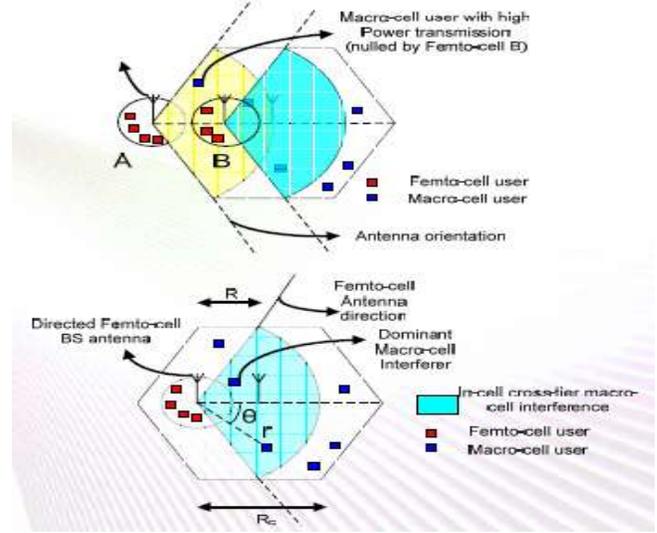

Fig. 11 Directional Antennas nullifying the cross tier interference

## 3. Simulation & Performance Evaluation

The basic assumptions for deployment of avoidance of the cross tier interference of femto cells in macro cells Networks are listed as follows;

Cognitive Power Control: Reduce Femto cell power by 25mw when femot cell user is closer (< 200m) to macrocell user, otherwise add 20mw.

Cross tier interference is modelled by Gaussian distribution between macro cell to femto cells and macro cell to macro cell.

The simulation result shows that the macro cell outage probability for $\chi_c$=12db using equation-16, due to cross tier interference from 24 surrounding inner femto cell BSs, surrounding macro cells BSs and their simultaneously growing femto cells BSs. It is found that macro cell outage gradually increased to 0.8 with simultaneously increasing 24 number of femto cell BSs deployment in neighbouring macro cells equipped with Omni-directional antenna as represented in Fig. 12. The macro cell outage is reduced to 0.4 and 0.2 at deploying 120 degree sectored antenna and 90 degree sectored antenna in simultaneously increasing 24 femto cell BSs. This improvement is optimized at macro cell outage to 0.25 and 0.13 by deploying 120 degree sectored antenna with Cognitive Power Control (CPC) and 90 degree sectored antenna with cognitive power control (CPC) in simultaneously increasing 24 femto cell BSs.

The simulation Specification is listed as follows;

TABLE I
SIMULATION PARAMETERS

| Element | Value |
| --- | --- |
| Radius of Macro cell (R) | 1000m |
| Radius of Femto cell (r) | 125*1.732 m |
| No. of Femto cell in a Macro cell (n) | 24 |
| Total Frequency allowed to use (F) | 1MHz |
| Frequency for Femto cell BS ($F_{femto}$) | 500 KHz @ 20.83 KHz |
| Frequency for Macro cell BS ($F_{macro}$) | 500 KHz |
| Standard Deviation (sigma) | 4 db |
| Alpha | 0.25 |
| Processing Gain | 256 db |
| Femto BS received power | 150 mw |
| Macro cell Target QoS or SIR | 12 db |
| Femto cell Target QoS or SIR | 14 db |
| Femto cell traffic Model | Poission's distribution |
| Max. Femto users at 1 BS at a time | 20 |
| Macro cell traffic Model | Gaussian distribution |
| Antenna Sectorization | 90 degree and 120 Degree. |
| Spectrum Sharing | Per Traffic Load balance. |

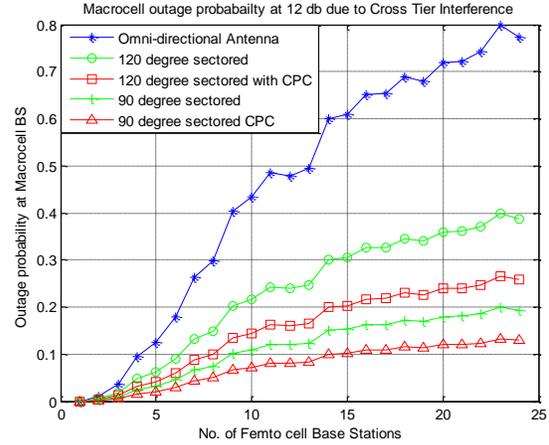

Fig.12 Macro cell Outage Probability

The following simulation result shows that the Femto cell outage probability at $\chi_f$=14db using equation-17 is increasing due to cross tier interference from increasing surrounding density of the interfering active femto cells users, different tiered femto cell BSs and surrounding femto cell BSs. It is observed that outage is optimised to 0.95 at 25 interfering macro cell users and found drastically increased to 1 at 50 interfering macro cell users as represented in Fig. 13. The femto cell outage is reduced to 0.5 and 0.2 at deploying 120 degree sectored antenna and 90 degree sectored antenna in femto cell BS as well as macro cell BSs. This is optimized for femto cell outage to 0.3 and 0.1 at deploying 120 degree sectored antenna with CPC and 90 degree sectored antenna with CPC in femto cell BS as well as macro cell BSs.

The given simulation refers the poission distribution of femto cell Traffic load at 24 femto cell BS within one macro cell. For each femto cell BS, 20 Femto cell users at one time is allowed with the total frequency of 20.83 KHz. The maximum spectrum utilization is found to be approximately 90% at the sixth, thirteenth and Twenty Femto cell BS because of higher traffic density as represented in Fig. 14.

The simulation also shows the spectrum sharing between femto cell and macro cell for 1MHz frequency where the spectrum is proportionately balanced and avoids the hot spot zones formation as well as cross tier interference as represented in Fig. 15.

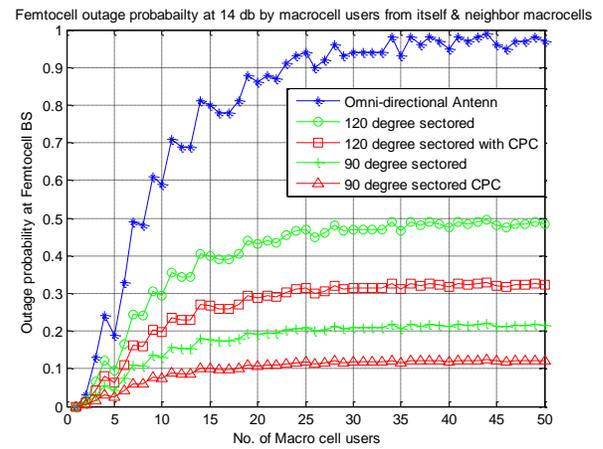

Fig.13 Femto cell Outage Probability

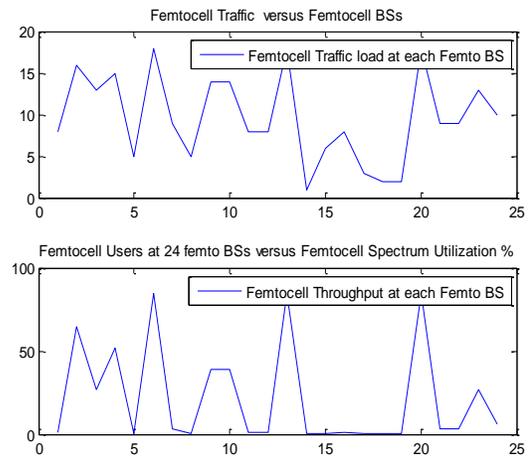

Fig. 14 Femto cell Traffic and Throughput

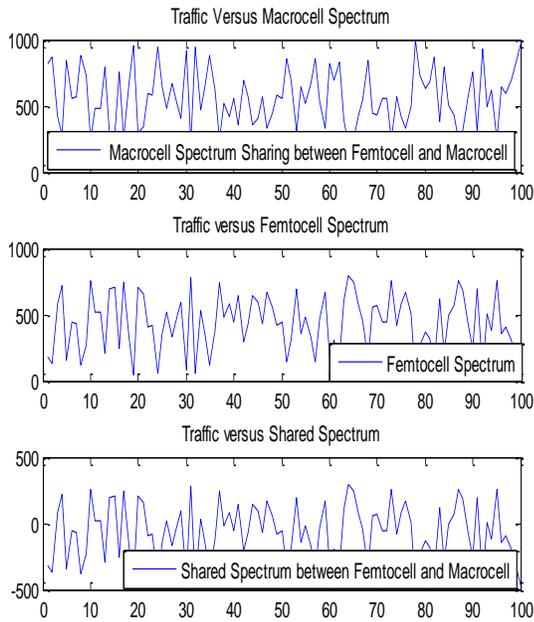

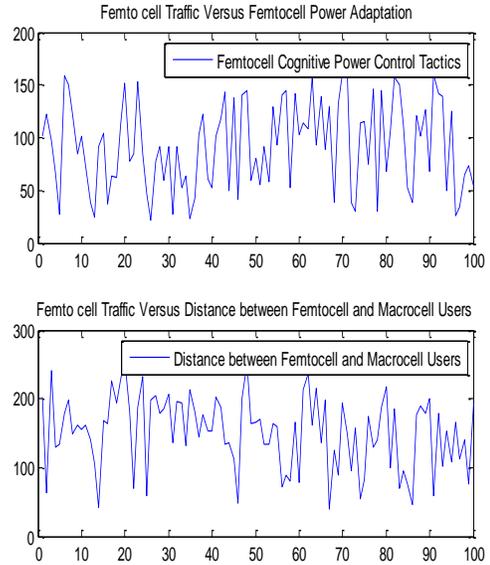

Fig. 15 Spectrum Sharing between Femto cell and Macro cell

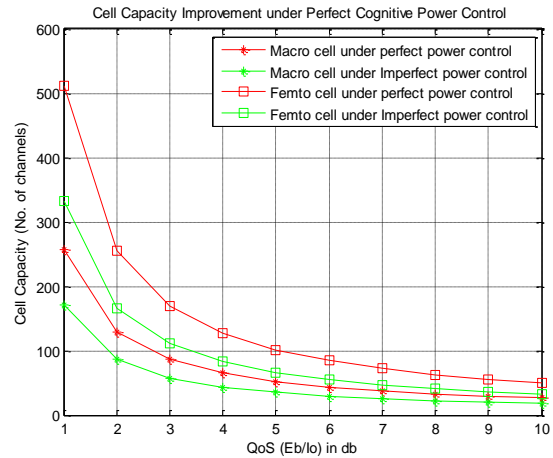

Fig. 17 Femto cell Cognitive Power Control Strategy with respect to Distance

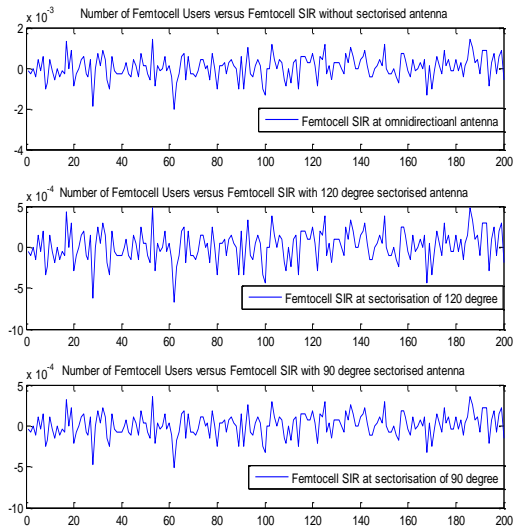

Fig. 16 Interference Mitigation by Antenna Sectorization

The simulation results show that the cross tier interference between femto cell and macro cell is proportionately mitigated by Directional Antennas sectored at 120 degree and 90 degree. From the simulation, the normal Gaussian cross tier interference is distributed to different regions by directional antennas such as 3 times by 120 degree sectorization and 4 times by 90 degree sectorization as represented in Fig. 16.

Fig. 18 Macro cell & Femto cell Capacity Improvement under Perfect Cognitive Power Control

The simulation results show that the Cognitive power control tactics implement the power adaptation based on the distance between femto cell users and macro cell users as shown in Fig. 17. As the distance between them becomes more closer than 200m then the femto cell user decreases power by 20mw to avoid the dead zone formation. Otherwise, femto cell user maintains its power by increasing 10mw.

The cell capacity for femto cell tiers and macro cell tiers has been significantly improved maintaining same QoS ($E_b/I_o$) deploying perfect cognitive power control strategy over imperfect power control as shown in Fig.18. The cell capacity is reduced to provide better Qos to the optimal number of channels. This becomes realistic as the power control error and source activity factor with imperfect power control are addressed by the perfect cognitive power control. From the simulation, it has been realized that maximum optimization of channel utilization is in femto cell rather than macro cell under perfect power control as compared to imperfect power control. The major reason is that femto cells are especially designed to provide the reliable link, higher coverage and better QoS to particular subscribers than others.

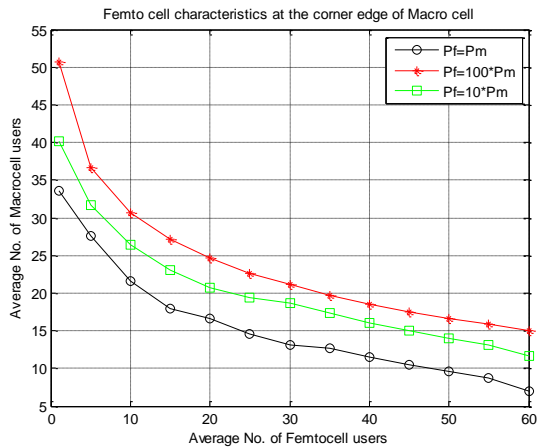

Fig. 19 Femto cell density characteristics

At the corner edge of the macro cell, femto cell BS provide significant power and coverage to femto cell users deploying multiple access by sectored beams as well as perfect cognitive power control to avoid interference from other tiered femto cell BS and macro cell users. In such scenario, when the density of femto cell users are highly populated then macro cell users are reduced by preventing from spectrum sharing and vice versa, to maintain threshold QoS as shown in above Fig. 19. Furthermore, higher power control factor defined from the proportion of the received femto cell and the received macro cell also increases the participation of average macro cell users maintaining same average number of femto cell users which can be adapted by cognitive power control strategy.

In brief, the optimal technique for the avoidance of cross tier interference between femto cell and macro cell networks is based on spectrum sharing, antenna sectorization and cognitive power control strategy. The simulation results also illustrate the outstanding performance of this optimization scheme. Initially, the DS-CDMA macro cell outage probability due to cross tier interference from surrounding femto cell BSs, another tier macro cell BS and its own macro cell users, is found significantly reduced deploying 90 degree sectored antenna than 120 degree sectored antenna and Omni-directional antenna with cognitive power control. Similarly, the femto cell outage probability due to cross tier interference from another tiered femto cell BSs, active femto cell users and intra femto cell BSs, is found significantly reduced deploying 90 degree sectored antenna than 120 degree sectored antenna and Omni-directional antenna with cognitive power control.

Furthermore, the Poission distribution of femto cell Traffic load at 24 femto cell BS considering 20 femto cell users at one time is allowed at each femtocell BS with the total frequency of 20.83 KHz. The maximum spectrum utilization is approximately 90% at the sixth, thirteenth and twenty femto cell BSs instantly. The spectrum sharing between femto cell and macro cell for 1MHz or 1000 KHz frequency is proportionally balanced and avoids the hot spot zones formation as well as cross tier interference. The antenna sectorization with 120 and 90 degree angle reduced cross tier interference by 3 and 4 times. Ultimately, the cognitive power control adapts distance based tactics to avoid dead zones as well as interference at the junction of directional antennas which increases channel capacity as well macro cell user's density. Thus, the integrated approach of the shared spectrum, sectored antenna and cognitive power control is the optimal solution for avoidance of cross tier interference between femto cell and macro cell Networks.

## 4. Conclusion

The cellular outage capacity depends upon the location of the user from base station, channel statistics and the co-channel interference. Similarly, the cell capacity is based on the power control error and multiple access interference in DS-CDMA. In addition, increasing QoS depends on the source activity factor and radically decreasing the processing gain as well as cell capacity. Consequently, femto cells are deployed in DS-CDMA macro cell to optimize outage and cell capacity with pre-margin QoS. In order to achieve this, the cross tier interference between femto cells users and macro cell users is evaded by the novel optimization approach of the shared spectrum for efficient traffic management, sectored antenna to optimize outage by mitigating cross interference and cognitive power control to optimize cell capacity and average density of macro cell users, with outstanding simulation results. Future work will focus on secured cross layer based cognitive radio deployment to optimize QoS in femto cell networks.

## References

[1] H.J. Claussen, L.Ho, L.G. Samuel, "An Overview of the Femtocell Concept", Bell Labs Technical Journal 13(1), 221–246, © 2008 Alcatel Lucent, Published online in Wiley Inter Science.

[2] O. Simeon, S. Erkip, S. Shitz, "Robust Transmission and Interference Management for Femtocells with Unreliable Network Access", IEEE JSAC, Vol. 28, No. 9, Dec 2010.

[3] V. Chandrasekhar, J.G. Andrews, "Uplink Capacity and Interference Avoidance for Two-Tier Femtocell Networks", IEEE Transactions on Wireless Communications, Vol. 8, No. 7, pp. 3498-3509, July 2009.

[4] S. Huan, K. Linling, L. Jianhua, "Interference avoidance in OFDMA-based femtocell network", IEEE Youth Conference on Information Computing and Telecommunication, 2009.

[5] V. Chandrasekhar, M. Kountouris, J. G. Andrews, "Coverage in Multi Antenna Two-Tier Networks", IEEE Transactions on Wireless Communications, Vol. 8, No. 10, pp. 5314-5327, October 2009.

[6] H. Claussen, "Performance of macro and co-channel femtocells in a hierarchical cell structure", IEEE PIMRC'07.

[7] S. Yeh, " WiMAX Femtocells : A Perspective on Network Architecture, Capacity, and Coverage", IEEE Communications Magazine, Oct. 2008.

[8] T. Nihtila, "Increasing femto cell throughput with HSDPA using higher order modulation", IEEE INCC, 2008.

[9] F. Chu, K. C. Chen, "Mitigation of Macro-Femto Co-channel Interference by Spatial Channel Separation", IEEE Vehicular Technology Conference, 2011.

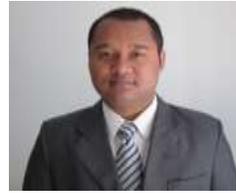

**Dr. Niraj Shakhakarmi** worked as a Doctoral Researcher since 2009-2011 in the US-Army Research Office (ARO) Center for Digital Battlefield Communications (CeBCom) Research, Department of Electrical and Computer Engineering, Prairie View A&M University. He received his B.E. degree in Computer Engineering in 2005 and M.Sc. in Information and Communication Engineering in 2007. His research interests are in the areas of Wireless Communications and Networks Security, Secured Position Location & Tracking (PL&T) of Malicious Radios, Cognitive Radio Networks, WCDMA/HSPA/LTE/WRAN Next Generations Wireless Networks, Satellite Networks and Digital Signal Processing, Wavelets Applications and Image/Colour Technology. He is a member of IEEE Communications Society, ISOC, IAENG and attended AMIE conference. He has published several WSEAS journals and IJCSI journals, along with WTS, Elsevier and ICSST conference papers. His several journals and conference papers are under review in IEEE journals. He is serving as reviewer for WSEAS, SAE, WTS and editorial board member for IJEECE, WASET, JCS and IJCN journals.